\documentclass[aps,prl,twocolumn,groupedaddress,showpacs]{revtex4-1}
\usepackage{graphicx}
\usepackage{amsmath}
\begin{document}

%\preprint{}

\title{Measurement of complex-valued spatial coherence \\and its application to diffractive imaging}


\author{Yifeng Shao}
\affiliation{Optics Research Group, Imaging Physics Department, Delft University of Technology, The Netherlands}
\author{Xingyuan Lu}
\affiliation{College of Physics, Optoelectronics and Energy and Collaborative Innovation Center of Suzhou Nano Science and Technology, Soochow University, China}
\affiliation{Key Lab of Advanced Optical Manufacturing Technologies of Jiangsu Province and Key Lab of Modern Optical Technologies of Education Ministry of China, Soochow University, China}
\author{Sander Konijnenberg}
\affiliation{Optics Research Group, Imaging Physics Department, Delft University of Technology, The Netherlands}
\author{Chengliang Zhao}
\email[zhaochengliang@suda.edu.cn]{}
\affiliation{College of Physics, Optoelectronics and Energy and Collaborative Innovation Center of Suzhou Nano Science and Technology, Soochow University, China}
\affiliation{Key Lab of Advanced Optical Manufacturing Technologies of Jiangsu Province and Key Lab of Modern Optical Technologies of Education Ministry of China, Soochow University, China}
\author{Yangjian Cai}
\affiliation{College of Physics, Optoelectronics and Energy and Collaborative Innovation Center of Suzhou Nano Science and Technology, Soochow University, China}
\affiliation{Key Lab of Advanced Optical Manufacturing Technologies of Jiangsu Province and Key Lab of Modern Optical Technologies of Education Ministry of China, Soochow University, China}
\author{H. Paul Urbach}
\affiliation{Optics Research Group, Imaging Physics Department, Delft University of Technology, The Netherlands}

\date{\today}

\begin{abstract}
The complete characterization of spatial coherence is difficult because the mutual coherence function is a complex-valued function of four independent variables. This difficulty limits the ability of controlling and optimizing spatial coherence in a broad range of key applications. Here we propose a method for measuring the complex-valued mutual coherence function, which does not require any prior knowledge and can be scaled to measure arbitrary coherence properties for any wavelength. We apply our method to retrieve a phase object illuminated by a partially coherent beam. This application is particularly useful for diffractive imaging of nanoscale structures in the X-ray or electron regime, where the illumination sources lack of spatial coherence.

\end{abstract}
%\pacs{}
%\keywords{}
\maketitle
%\section{}
Spatial coherence is among the fundamental properties of light. It describes the statistical correlation between fields at different positions. However, the measurement of spatial coherence is remarkably challenging, which limits the control and the optimization of spatial coherence in a broad range of key applications such as beam shaping using spatial coherence as an extra degree of freedom \cite{Cai:2014}, optical communication through a turbulent atmosphere \cite{Gbur:2014}, illumination for advanced imaging systems (e.g. optical lithography) \cite{Lai:2009} and superresolution imaging \cite{Douglass:2016}. The spatial coherence of a beam is described by the complexed-valued 4-dimensional mutual coherence function (MCF). The process of measuring the MCF is extremely time consuming, especially for methods measuring the intensity-intensity correlation \cite{Cai:2014,Chen:2014} or interference \cite{Partanen:2014} at all pairs of points. Other interference based methods measure the fringe visibility in a single direction \cite{Pfeiffer:2005,Divitt:2015, Morrill:2016} or multiple directions \cite{Marathe:2014, Shi:2014}. Spatial coherence can alternatively be determined using the so-called phase space methods \cite{Tran:2007,Waller:2012,Wood:2014,Sharma:2016}. However, all of the reported methods either assume the MCF to be shift-invariant \cite{Pfeiffer:2005,Divitt:2015, Morrill:2016,Marathe:2014, Shi:2014} or lack the ability to reveal both the amplitude and the phase of the MCF \cite{Cai:2014,Chen:2014,Marathe:2014, Shi:2014,Tran:2007,Waller:2012,Wood:2014,Sharma:2016}. Moreover, a preferred method should not rely on any prior knowledge, for example an analytical model \cite{Liu:2017}, so that arbitrary coherence properties can be measured. 

Diffractive imaging is an essential application of spatial coherence, which aims to reconstruct the nanoscale structures of an object from its diffracted far field intensity. At short wavelengths, e.g. in the X-ray and electron regime, the illumination sources lack spatial coherence. This severely degrades the result of imaging. Experimentally,  almost completely coherent bream can be obtained by using coherence filtering which requires a small aperture and a long propagation distance in vacuum. In order to utilize the partially coherent beam for illumination, the widely used iterative algorithms have been modified to take into account the MCF instead of the electromagnetic field. However, this modification is restricted by the accuracy of the approximation used to propagate the MCF, which either assumes a shift-invariant MCF and describes the recorded intensity as a convolution \cite{Clark:2012, Burdet:2015} or is based on mode decomposition of the MCF \cite{Whitehead:2009, Thibault:2013}. To our knowledge, none of the non-iterative algorithms, e.g. \cite{McNulty:1992, Marchesini:2008}, has been adapted to utilize partially coherent beam, despite the obvious advantages of instant and unambiguous reconstruction of the object. 

In this paper, we present a method for measuring the complete MCF of an arbitrary partially cohernet light beam. We let this beam propagate through a window with finite size and we measure the diffracted far field intensities. By introducing a perturbation at a particular point, we can retrieve a 2-dimensional slice of the 4-dimensional MCF linked to the location of the perturbation point. Hence the complete MCF can be obtained by a 2-dimensional scanning procedure.

Our method can be turned into a diffractive imaging method by superposing the window with a transmissive object. In this case we retrieve the slice of the MCF in the plane right behind the object. The use of partially coherent illumination leads to a modulation effect of the object. After having calibrated and compensated for this modulation, we can then retrieve the object alone. Here we investigate the performance of our method using beams with various types of correlation sturand degrees of coherence via a proof-of-principle experiment at visible wavelength. Our method uses a lensless configuration and is therefore wavelength independent.

\begin{figure}[t]
%\fbox
\includegraphics[width=\linewidth]{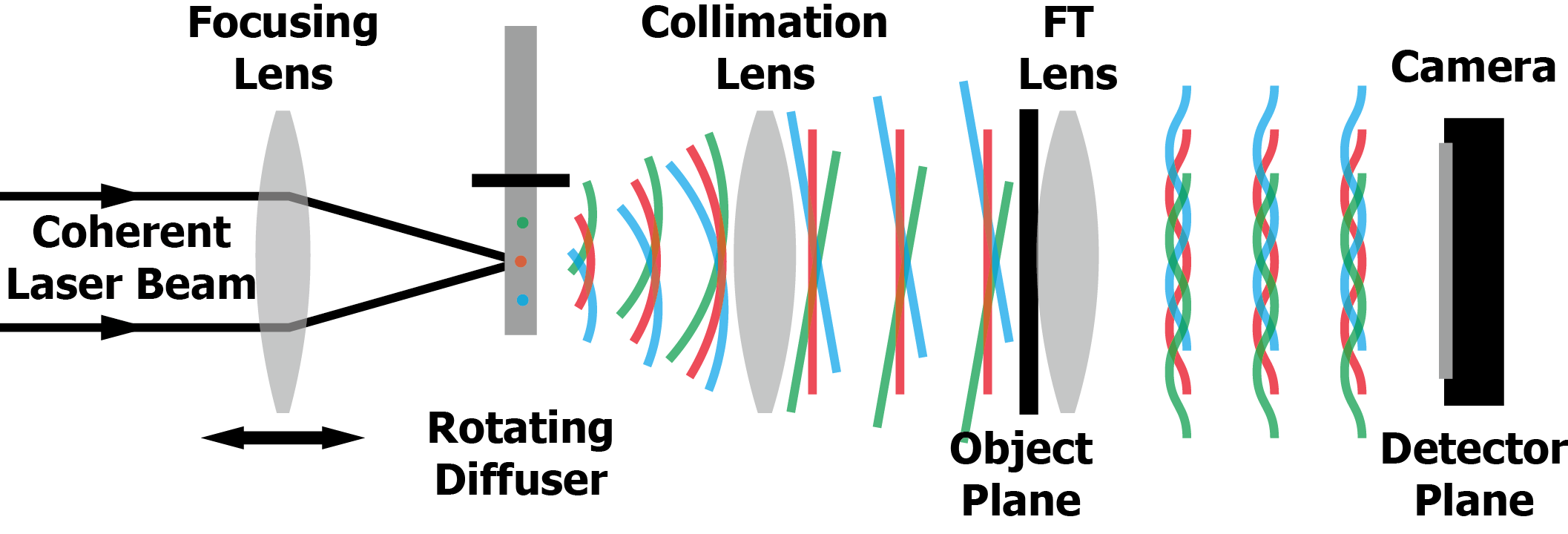}
\caption{Experimental setup. The light scattered by the rotating diffuser can be regarded as emitted by a collection of independent point sources, which generates the partially coherent beam for illumination. The field at the detector plane is the Fourier transform of the field transmitted by the object.}
\label{fig:figure1}
\end{figure}

The experimental setup is illustrated in Fig.~\ref{fig:figure1}. The MCF at the rotating diffuser plane is given by
\begin{equation}
J_S(\boldsymbol{r}_1,\boldsymbol{r}_2) = 
\delta(\boldsymbol{r}_1,\boldsymbol{r}_2)
S(\boldsymbol{r}_1)S(\boldsymbol{r}_2)^*,
\end{equation}
where $\boldsymbol{r}$ denotes the coordinate in this plane the subscript of $\boldsymbol{r}$ distinguishes two points. $J_S(\boldsymbol{r}_1,\boldsymbol{r}_2)$ represents a collection of independent point sources. This incoherent source generates a partially cohernet beam and the MCF in the object plane is 
\begin{multline}\label{eq:MCF}
J(\boldsymbol{\rho}_1,\boldsymbol{\rho}_2) = 
\iiiint J_S(\boldsymbol{r}_1,\boldsymbol{r}_2)\cdots \\
\times\exp\left[-i2\pi(\boldsymbol{\rho}_1\boldsymbol{r}_1
-\boldsymbol{\rho}_2\boldsymbol{r}_2)\right]
\mathrm{d}^2\boldsymbol{r}_1\mathrm{d}^2\boldsymbol{r}_2,
\end{multline}
where $\boldsymbol{\rho}$ denotes the object plane coordinate. Eq.~(\ref{eq:MCF}) indicates that the MCF of the partially coherent beam generated in this way is essentially complex-valued.

The intensity distribution of the incoherent source (the focal spot) $|S(\boldsymbol{r})|$ determines the properties of the partially coherent beam. In the experiment, we can obtain two different focal spot intensity distributions and generate two differently correlated beams: the Gaussian correlated beam and the Gaussian-Airy correlated beam. The difference is that the coherent laser beam is truncated by the focusing lens in the later case but not in the former case. To vary the degree of coherence, we change the size of the focal spot by translating back and forth the focusing lens while keeping the type of correlation. 

Let the generated partially coherent beam propagate through a phase object. We create this object by programming a square of $240\times240$ pixels on the phase-only spatial light modulator (SLM). The object is binary, whose value can only be $0.1\pi$ or $0.9\pi$, in the shape of a panda. The MCF of the partially coherent field right behind the object is
\begin{equation}
J'(\boldsymbol{\rho}_1,\boldsymbol{\rho}_2) = J(\boldsymbol{\rho}_1,\boldsymbol{\rho}_2)
O(\boldsymbol{\rho}_1)O(\boldsymbol{\rho}_2)^*. 
\end{equation}
The field at the object plane is related to the field at the detector plane by Fourier transform (FT). This could be achieved through a Fresnel or a Fraunhofer propagation in free space, e.g. in the X-ray and electron regime, or by using a FT lens as we did for visible wavelengths. The diffracted far field intensity of the partially coherent field in the object plane can be expressed by:
\begin{multline}\label{eq:Intensity0}
I_0(\boldsymbol{r}') = \iiiint J'(\boldsymbol{\rho}_1,\boldsymbol{\rho}_2)\cdots\\
\times\exp\left[-i2\pi\boldsymbol{r}'(\boldsymbol{\rho}_1-\boldsymbol{\rho}_2)\right]
\mathrm{d}^2\boldsymbol{\rho}_1\mathrm{d}^2\boldsymbol{\rho}_2,
\end{multline}
where $\boldsymbol{r}'$ denotes the detector plane coordinate. Here we assume that the object has been truncated by the window. Using a constant object is thus equivalent to applying a window to the incident partially coherent beam.

We introduce a perturbation $\gamma\delta(\boldsymbol{\rho}-\boldsymbol{\rho}_0)$ at point $\boldsymbol{\rho}=\boldsymbol{\rho}_0$, where $\gamma$ is a complex-valued constant. The intensity distribution can then be written as:
\begin{equation}\label{eq:Intensity}
\begin{aligned}
&I(\boldsymbol{r}) = I_0(\boldsymbol{r}) +|\gamma|^2J(\boldsymbol{\rho}_0,\boldsymbol{\rho}_0)\\
&+\begin{aligned}
\iint J(\boldsymbol{\rho}_1,\boldsymbol{\rho}_0)
O(\boldsymbol{\rho}_1)[\gamma O(\boldsymbol{\rho}_0)]^*\cdots\\
\times\exp\left[-i2\pi\boldsymbol{r}(\boldsymbol{\rho}_1-\boldsymbol{\rho}_0)\right]
\mathrm{d}^2\boldsymbol{\rho}_1\end{aligned}\\
&+\begin{aligned}
\iint J(\boldsymbol{\rho}_0,\boldsymbol{\rho}_2)
[\gamma O(\boldsymbol{\rho}_0)]O(\boldsymbol{\rho}_2)^*\cdots\\
\times\exp\left[-i2\pi\boldsymbol{r}(\boldsymbol{\rho}_0-\boldsymbol{\rho}_2)\right]
\mathrm{d}^2\boldsymbol{\rho}_2\end{aligned}
\end{aligned}.
\end{equation}
Ideally, the perturbation should be only at a point, so that it can be approximated by a delta function. In the experiment, the perturbed region is a square of $20\times20$ pixels. When the perturbation point is located inside the object, we need to assign a complex-valued constant to the values of pixels in the perturbed region. When the perturbation point is located outside the object, we only need to switch the state of the pixels in the perturbed region on and off. 

The inverse FT of this intensity distribution yields:
\begin{equation}\label{eq:IntensityFT}
\begin{aligned}
\tilde{I}(\boldsymbol{r}')(\boldsymbol{\rho}) &= 
\tilde{I}_0(\boldsymbol{r}')(\boldsymbol{\rho})+
|\gamma|^2 J(\boldsymbol{\rho}_0,\boldsymbol{\rho}_0)
\delta(\boldsymbol{\rho}) \\
&+J(\boldsymbol{\rho}_0+\boldsymbol{\rho},\boldsymbol{\rho}_0)
O(\boldsymbol{\rho}_0+\boldsymbol{\rho})[\gamma O(\boldsymbol{\rho}_0)]^*\\
&+ J(\boldsymbol{\rho}_0,\boldsymbol{\rho}_0-\boldsymbol{\rho})
[\gamma O(\boldsymbol{\rho}_0)]O(\boldsymbol{\rho}_0-\boldsymbol{\rho})^*
\end{aligned},
\end{equation}
where $\tilde{\cdot}$ denotes the operation of inverse FT. We can observe two cross terms occurring in Eq.~(\ref{eq:IntensityFT}): one is proportional to the object $O(\boldsymbol{\rho})$, referred to as the ``reconstructed object'', and the other is proportional to the complex conjugated of the object rotated by 180 degrees $O(-\boldsymbol{\rho})^*$, referred to as the ``twin image''.

\begin{figure}[t]
%\fbox
\includegraphics[width=\linewidth]{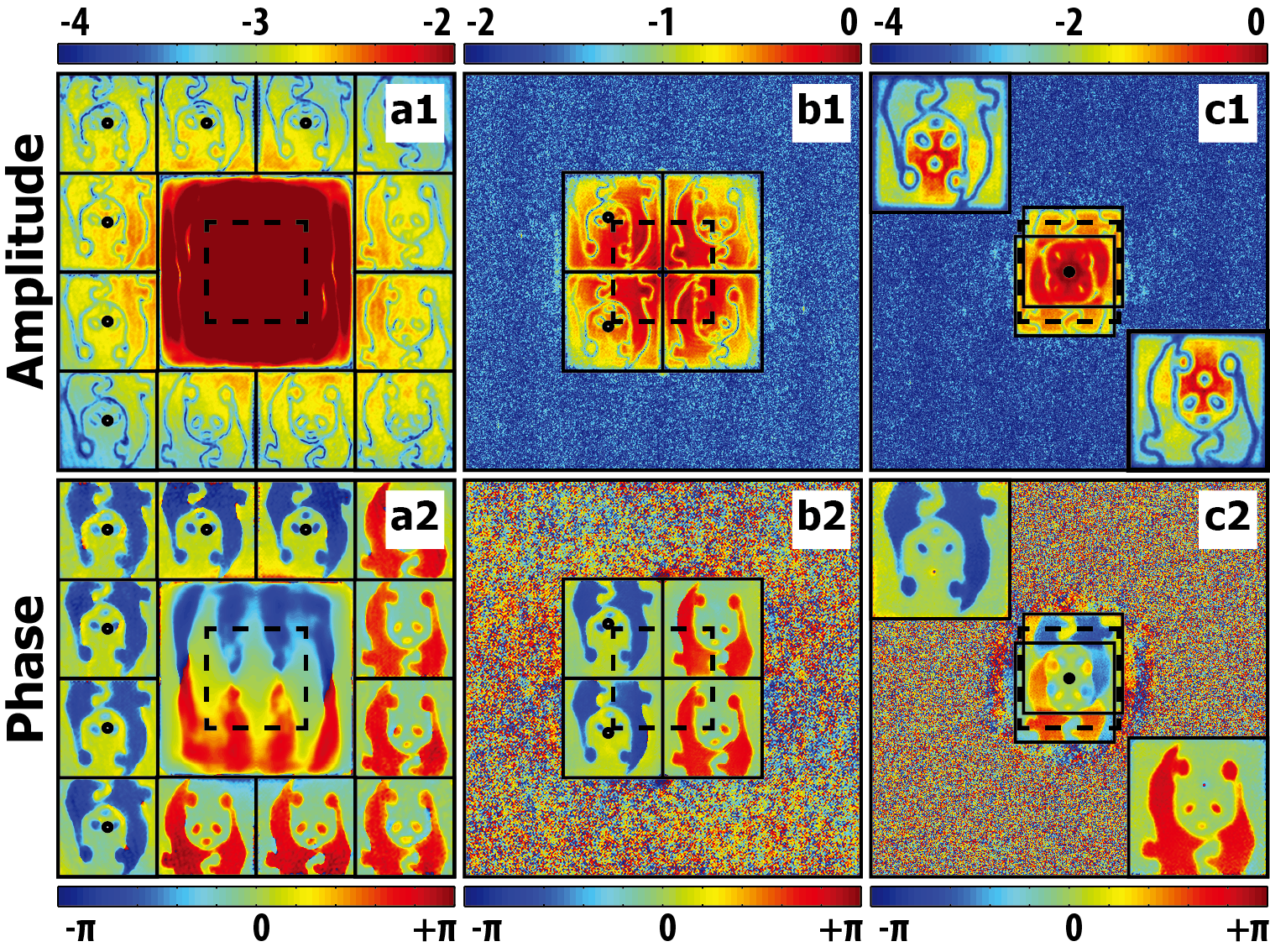}
\caption{Experiment results of different overlaps between the two cross terms and the quadratic term using Gaussian correlated illumination. (a1,2): The single measurement case using 6 perturbation points. (b1,2): The double measurement case using 2 perturbation points. (c1,2): The triple measurement case using only 1 perturbation point. The dashed square shows the finite size of the window and the dots show the locations of perturbation points. The red and blue panda are the "reconstructed object" and the "twin image" respectively. }
\label{fig:figure2}
\end{figure}

We need to separate and retrieve both two cross terms. In Eq.~(\ref{eq:IntensityFT}), the quadratic term $\tilde{I}_0(\boldsymbol{r})(\boldsymbol{\rho})$ is the inverse FT of the intensity distribution without perturbation and is centered at the origin. The two cross terms are located symmetrically on opposite sides of the origin. In our method, the location of perturbation point can be chosen freely in the object plane. This freedom of choice influences the amount of overlap between the quadratic term and the two cross terms. Depending on the overlapping, three different situations occur, which differ in the number of required measurement and the treatments for computationally retrieving the two cross terms. We illustrate different overlap situations in Fig.~\ref{fig:figure2} using a Gaussian correlated beam for illumination.

When the two cross terms do not overlap with the quadratic term as shown Fig.~\ref{fig:figure2}(a), our method requires only one measurement. Perturbation points at different locations can be introduced at the same time, provided that the generated pairs of cross terms do not overlap with each other. In this case, we introduce six perturbation points which generate six pairs of cross terms filling the area surrounding the quadratic term.

When the two cross terms overlap with the quadratic term but not with each other as shown in Fig.~\ref{fig:figure2}.(b), our method requires two measurements. The extra measurement is the intensity distribution without perturbation $I_0(\mathbf{r})$. It will be used to eliminate the quadratic term. In this case we introduce two perturbation points at the two left corner of the object. The two pairs of cross terms are generated which occupy the area of the quadratic term. 

When the two cross terms overlap with the quadratic term and with each other as shown in Fig.~\ref{fig:figure2}.(c), our method requires three measurements. Each measurement corresponding to a different values of $\gamma$ at the same perturbation point. Note that the intensity measurement without perturbation correspond to the case of $\gamma = 0$.
We first eliminate the quadratic term using $I_0(\mathbf{r})$, then solve for both two cross terms using the two remaining linear equations. In this case, only one perturbation perturbation point can be introduced. 

The "reconstructed object" is actually the product of $O(\boldsymbol{\rho})$ and a 2-dimensional slice of the 4-dimensional MCF $J(\boldsymbol{\rho},\boldsymbol{\rho}_0)$, which describes the correlation between the field in the object plane and the field at the location of perturbation point $\boldsymbol{\rho} = \boldsymbol{\rho}_0$. To measure the complete MCF, we need to perform a 2-dimensional scanning of $\boldsymbol{\rho}_0$ in the object plane. 

As can be seen in Eq.~(\ref{eq:IntensityFT}), the product $J(\boldsymbol{\rho},\boldsymbol{\rho}_0)O(\boldsymbol{\rho})$ has been shifted by $-\boldsymbol{\rho}_0$. Thus it occurs at different positions according to the location of the perturbation point. In Fig.~\ref{fig:figure2} we use a Gaussian correlated beam whose MCF is shift-invariant. Because the maximum of all slice of this MCF will always stay at the origin, the amplitude of the ''reconstructed object'' is modulated by different part of the same Gaussian distribution $J(\boldsymbol{\rho})$ as we vary the location of the perturbation point, while the phase is not modulated. Fig.~\ref{fig:figure2} shows that the amplitude is modulated by a stronger part of $J(\boldsymbol{\rho})$ for smaller $\boldsymbol{\rho}_0$ and by a weaker part of $J(\boldsymbol{\rho})$ for larger $\boldsymbol{\rho}_0$.

\begin{figure}[b]
%\fbox
\includegraphics[width=\linewidth]{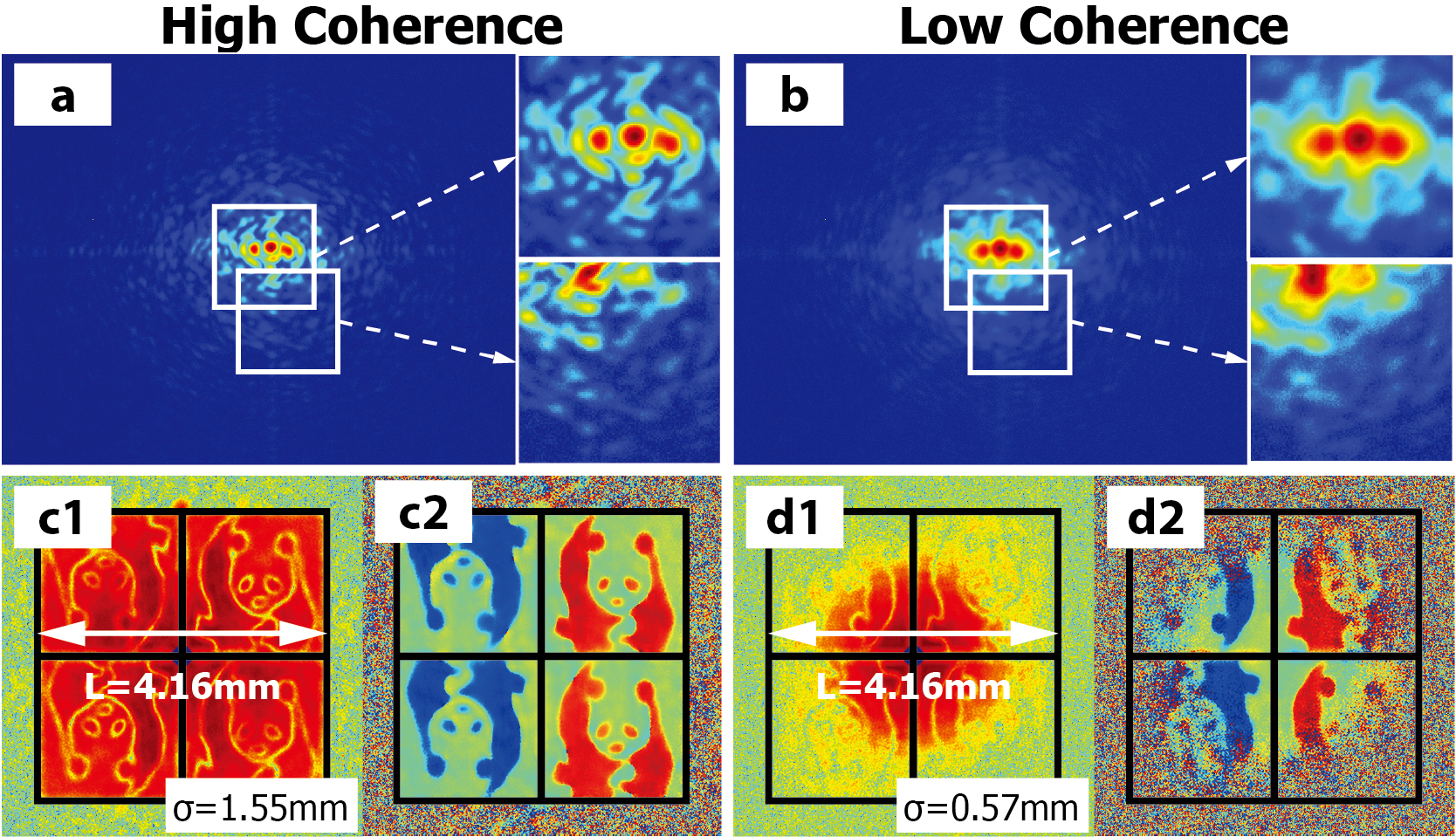}
\caption{Experiment results using Gaussian correlated illumination for different degrees of coherence. (a,b): The intensity measurements of the diffracted far field. The inserted plots are detailed views of the region of interest in the intensity measurements. (c,d): The amplitude (1) and the phase (2) of the MCF of a Gaussian correlated beam transmitted through a phase object. $L$ denotes twice the size of the object and $\sigma$ denotes the waist of the Gaussian distribution used to fit the amplitude profile.}
\label{fig:figure3}
\end{figure}

We compare the experiment results of double measurement case using Gaussian correlated beam with two different degree of coherence Fig.~\ref{fig:figure3}. Two perturbation points are introduced at the two left corners of the object. In Fig.~\ref{fig:figure3}(c,d), the two "reconstructed objects" and the two "twin images" are located in the right and left half, and are represented by the red and blue panda respectively. We fit the profile of the amplitude to the Gaussian distribution to find its waist $\sigma$. The result suggests that the modulation of this amplitude indeed obeys the Gaussian distribution. We observe that the degree of coherence determines the waist as well as the blurry of the intensity measurement in Fig.~\ref{fig:figure3}(a,b). 

Fig.~\ref{fig:figure3} shows that random fluctuation does not destroy the phase correlation of a partially coherent field. The resolution of the phase is independent of the degree of coherence. However, whether we can retrieve the phase depends on whether the modulated amplitude can overcome the noise. For diffractive imaging, this mechanism determines the field-of-view. The center of the field-of-view will always be at the origin where the amplitude correlation is the maximum. Fixing the noise level in the measurement, the field-of-view is larger for higher coherence and smaller for lower coherence.

\begin{figure}[t]
%\fbox
\includegraphics[width=\linewidth]{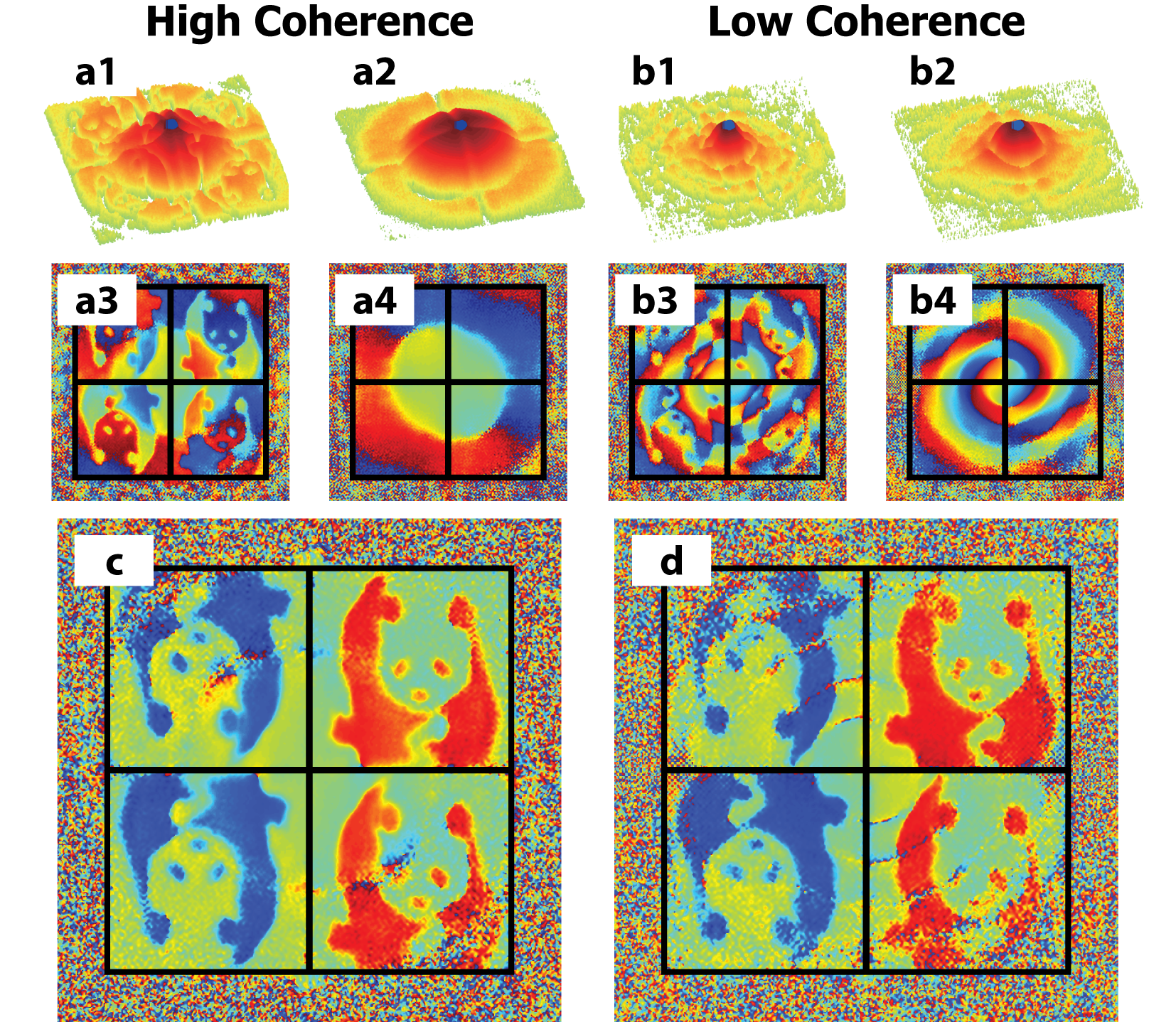}
\caption{Experiment results using Gaussian-Airy correlated beam for different degrees of coherence. (a1,3 and b1,3): The amplitude and the phase of the MCF of a partially coherent beam transmitted through an object. (a2,4 and b2,4): The amplitude and phase of the MCF of illumination beam. (c and d): The results of diffractive imaging after the modulation has been compensated for.}
\label{fig:figure4}
\end{figure}

We have demonstrated that we can retrieve a slice of the MCF of a partially coherent beam transmitted through a phase object. The information of the object is contained in both cross terms. The use of partially coherent beam leads to a modulation effect. In Fig.~\ref{fig:figure3}.(c2,d2), the phase is not modulated due to the use of a simple Gaussian correlated beam. However, using more complicated beams, e.g. the Gaussian-Airy correlated beam, the modulation of phase becomes visible as shown in Fig.~\ref{fig:figure4}.(a3,b3). Therefore, to obtain the object alone, we need to calibrate and compensate for this modulation. 

The calibration of modulation requires the perturbation points to be located at exactly the same positions in the presence of the object. We plot the amplitude and the phase of the slice of MCF of the incident partially coherent beam in Fig.~\ref{fig:figure4}.(a2,b2) and Fig.~\ref{fig:figure4}.(a4,b4) respectively. The amplitude exhibits an Airy pattern, while its phase jumps between $0$ and $\pi$ at the locations where this Airy pattern reaches zero, which means that the correlation of two fields changes its sign. To compensate for the modulation, we divide the retrieved slice of MCF in the presence of the object by the calibrated slice of MCF. We demonstrate the diffractive imaging results in Figure.~\ref{fig:figure4}.(c.d) for two degrees of coherence. We found that provided sufficient signal to noise ratio, the results can be independent of the properties of the incident partially coherent beam. 

To conclude, we propose and experimentally demonstrate an accurate yet robust method for measuring the complete complex-valued MCF as well as for diffractive imaging using a partially coherent beam for illumination. The resolution of the retrieved MCF is determined by the size of the perturbed region, while the field-of-view is determined by the coherence properties and the noise in the measurements. In the case of complete coherent illumination, our method is identical to three well-established holographic methods \cite{McNulty:1992,Codona:2013,Lai:2000} in each of the three situations, and hence shares identical limitations of the resolution and the field-of-view. Our method provides an alternative for rigorously studying the spatial coherence effect. It also opens the opportunities for developing new algorithms to make the imaging at x-ray and electron regime faster and simpler.

\begin{acknowledgments}
This work is part of the research programme “Novel design shapes for complex optical systems” with project number 12798, which is (partly) financed by the Netherlands Organisation for Scientific Research (NWO). This work is also supported by the National Natural Science Foundation of China (11374222), National Natural Science Fund for Distinguished Young Scholars (11525418), the Qing Lan Project of Jiangsu Province, and the sponsorship of Jiangsu Overseas Research \& Training program for University Prominent Young \& Middle-aged Teachers and Presidents.
%put your acknowledgments here.
\end{acknowledgments}


%\bibliography{ref}{}

\begin{thebibliography}{24}%
\makeatletter
\providecommand \@ifxundefined [1]{%
 \@ifx{#1\undefined}
}%
\providecommand \@ifnum [1]{%
 \ifnum #1\expandafter \@firstoftwo
 \else \expandafter \@secondoftwo
 \fi
}%
\providecommand \@ifx [1]{%
 \ifx #1\expandafter \@firstoftwo
 \else \expandafter \@secondoftwo
 \fi
}%
\providecommand \natexlab [1]{#1}%
\providecommand \enquote  [1]{``#1''}%
\providecommand \bibnamefont  [1]{#1}%
\providecommand \bibfnamefont [1]{#1}%
\providecommand \citenamefont [1]{#1}%
\providecommand \href@noop [0]{\@secondoftwo}%
\providecommand \href [0]{\begingroup \@sanitize@url \@href}%
\providecommand \@href[1]{\@@startlink{#1}\@@href}%
\providecommand \@@href[1]{\endgroup#1\@@endlink}%
\providecommand \@sanitize@url [0]{\catcode `\\12\catcode `\$12\catcode
  `\&12\catcode `\#12\catcode `\^12\catcode `\_12\catcode `\%12\relax}%
\providecommand \@@startlink[1]{}%
\providecommand \@@endlink[0]{}%
\providecommand \url  [0]{\begingroup\@sanitize@url \@url }%
\providecommand \@url [1]{\endgroup\@href {#1}{\urlprefix }}%
\providecommand \urlprefix  [0]{URL }%
\providecommand \Eprint [0]{\href }%
\providecommand \doibase [0]{http://dx.doi.org/}%
\providecommand \selectlanguage [0]{\@gobble}%
\providecommand \bibinfo  [0]{\@secondoftwo}%
\providecommand \bibfield  [0]{\@secondoftwo}%
\providecommand \translation [1]{[#1]}%
\providecommand \BibitemOpen [0]{}%
\providecommand \bibitemStop [0]{}%
\providecommand \bibitemNoStop [0]{.\EOS\space}%
\providecommand \EOS [0]{\spacefactor3000\relax}%
\providecommand \BibitemShut  [1]{\csname bibitem#1\endcsname}%
\let\auto@bib@innerbib\@empty
%</preamble>
\bibitem [{\citenamefont {Cai}\ \emph {et~al.}(2014)\citenamefont {Cai},
  \citenamefont {Chen},\ and\ \citenamefont {Wang}}]{Cai:2014}%
  \BibitemOpen
  \bibfield  {author} {\bibinfo {author} {\bibfnamefont {Y.}~\bibnamefont
  {Cai}}, \bibinfo {author} {\bibfnamefont {Y.}~\bibnamefont {Chen}}, \ and\
  \bibinfo {author} {\bibfnamefont {F.}~\bibnamefont {Wang}},\ }\href {\doibase
  10.1364/JOSAA.31.002083} {\bibfield  {journal} {\bibinfo  {journal} {Journal
  of the Optical Society of America A}\ }\textbf {\bibinfo {volume} {31}},\
  \bibinfo {pages} {2083} (\bibinfo {year} {2014})}\BibitemShut {NoStop}%
\bibitem [{\citenamefont {Gbur}(2014)}]{Gbur:2014}%
  \BibitemOpen
  \bibfield  {author} {\bibinfo {author} {\bibfnamefont {G.}~\bibnamefont
  {Gbur}},\ }\href@noop {} {\bibfield  {journal} {\bibinfo  {journal} {Journal
  of the Optical Society of America A}\ }\textbf {\bibinfo {volume} {31}},\
  \bibinfo {pages} {2038} (\bibinfo {year} {2014})}\BibitemShut {NoStop}%
\bibitem [{\citenamefont {Lai}\ \emph {et~al.}(2009)\citenamefont {Lai},
  \citenamefont {Rosenbluth}, \citenamefont {Bagheri}, \citenamefont
  {Hoffnagle}, \citenamefont {Tian}, \citenamefont {Melville}, \citenamefont
  {Tirapu-Azpiroz}, \citenamefont {Fakhry}, \citenamefont {Kim}, \citenamefont
  {Halle}, \citenamefont {McIntyre}, \citenamefont {Wagner}, \citenamefont
  {Burr}, \citenamefont {Burkhardt}, \citenamefont {Corliss}, \citenamefont
  {Gallagher}, \citenamefont {Faure}, \citenamefont {Hibbs}, \citenamefont
  {Flagello}, \citenamefont {Zimmermann}, \citenamefont {Kneer}, \citenamefont
  {Rohmund}, \citenamefont {Hartung}, \citenamefont {Hennerkes}, \citenamefont
  {Maul}, \citenamefont {Kazinczi}, \citenamefont {Engelen}, \citenamefont
  {Carpaij}, \citenamefont {Groenendijk}, \citenamefont {Hageman},\ and\
  \citenamefont {Russ}}]{Lai:2009}%
  \BibitemOpen
  \bibfield  {author} {\bibinfo {author} {\bibfnamefont {K.}~\bibnamefont
  {Lai}}, \bibinfo {author} {\bibfnamefont {A.~E.}\ \bibnamefont {Rosenbluth}},
  \bibinfo {author} {\bibfnamefont {S.}~\bibnamefont {Bagheri}}, \bibinfo
  {author} {\bibfnamefont {J.}~\bibnamefont {Hoffnagle}}, \bibinfo {author}
  {\bibfnamefont {K.}~\bibnamefont {Tian}}, \bibinfo {author} {\bibfnamefont
  {D.}~\bibnamefont {Melville}}, \bibinfo {author} {\bibfnamefont
  {J.}~\bibnamefont {Tirapu-Azpiroz}}, \bibinfo {author} {\bibfnamefont
  {M.}~\bibnamefont {Fakhry}}, \bibinfo {author} {\bibfnamefont
  {Y.}~\bibnamefont {Kim}}, \bibinfo {author} {\bibfnamefont {S.}~\bibnamefont
  {Halle}}, \bibinfo {author} {\bibfnamefont {G.}~\bibnamefont {McIntyre}},
  \bibinfo {author} {\bibfnamefont {A.}~\bibnamefont {Wagner}}, \bibinfo
  {author} {\bibfnamefont {G.}~\bibnamefont {Burr}}, \bibinfo {author}
  {\bibfnamefont {M.}~\bibnamefont {Burkhardt}}, \bibinfo {author}
  {\bibfnamefont {D.}~\bibnamefont {Corliss}}, \bibinfo {author} {\bibfnamefont
  {E.}~\bibnamefont {Gallagher}}, \bibinfo {author} {\bibfnamefont
  {T.}~\bibnamefont {Faure}}, \bibinfo {author} {\bibfnamefont
  {M.}~\bibnamefont {Hibbs}}, \bibinfo {author} {\bibfnamefont
  {D.}~\bibnamefont {Flagello}}, \bibinfo {author} {\bibfnamefont
  {J.}~\bibnamefont {Zimmermann}}, \bibinfo {author} {\bibfnamefont
  {B.}~\bibnamefont {Kneer}}, \bibinfo {author} {\bibfnamefont
  {F.}~\bibnamefont {Rohmund}}, \bibinfo {author} {\bibfnamefont
  {F.}~\bibnamefont {Hartung}}, \bibinfo {author} {\bibfnamefont
  {C.}~\bibnamefont {Hennerkes}}, \bibinfo {author} {\bibfnamefont
  {M.}~\bibnamefont {Maul}}, \bibinfo {author} {\bibfnamefont {R.}~\bibnamefont
  {Kazinczi}}, \bibinfo {author} {\bibfnamefont {A.}~\bibnamefont {Engelen}},
  \bibinfo {author} {\bibfnamefont {R.}~\bibnamefont {Carpaij}}, \bibinfo
  {author} {\bibfnamefont {R.}~\bibnamefont {Groenendijk}}, \bibinfo {author}
  {\bibfnamefont {J.}~\bibnamefont {Hageman}}, \ and\ \bibinfo {author}
  {\bibfnamefont {C.}~\bibnamefont {Russ}},\ }in\ \href@noop {} {\emph
  {\bibinfo {booktitle} {SPIE Advanced Lithography}}},\ Vol.\ \bibinfo {volume}
  {7274}\ (\bibinfo  {publisher} {International Society for Optics and
  Photonics},\ \bibinfo {year} {2009})\ pp.\ \bibinfo {pages}
  {72740A--72740A--12}\BibitemShut {NoStop}%
\bibitem [{\citenamefont {Douglass}\ \emph {et~al.}(2016)\citenamefont
  {Douglass}, \citenamefont {Sieben}, \citenamefont {Archetti}, \citenamefont
  {Lambert},\ and\ \citenamefont {Manley}}]{Douglass:2016}%
  \BibitemOpen
  \bibfield  {author} {\bibinfo {author} {\bibfnamefont {K.~M.}\ \bibnamefont
  {Douglass}}, \bibinfo {author} {\bibfnamefont {C.}~\bibnamefont {Sieben}},
  \bibinfo {author} {\bibfnamefont {A.}~\bibnamefont {Archetti}}, \bibinfo
  {author} {\bibfnamefont {A.}~\bibnamefont {Lambert}}, \ and\ \bibinfo
  {author} {\bibfnamefont {S.}~\bibnamefont {Manley}},\ }\href@noop {}
  {\bibfield  {journal} {\bibinfo  {journal} {Nat Photon}\ }\textbf {\bibinfo
  {volume} {10}},\ \bibinfo {pages} {705} (\bibinfo {year} {2016})}\BibitemShut
  {NoStop}%
\bibitem [{\citenamefont {Chen}\ \emph {et~al.}(2014)\citenamefont {Chen},
  \citenamefont {Wang}, \citenamefont {Liu}, \citenamefont {Zhao},
  \citenamefont {Cai},\ and\ \citenamefont {Korotkova}}]{Chen:2014}%
  \BibitemOpen
  \bibfield  {author} {\bibinfo {author} {\bibfnamefont {Y.}~\bibnamefont
  {Chen}}, \bibinfo {author} {\bibfnamefont {F.}~\bibnamefont {Wang}}, \bibinfo
  {author} {\bibfnamefont {L.}~\bibnamefont {Liu}}, \bibinfo {author}
  {\bibfnamefont {C.}~\bibnamefont {Zhao}}, \bibinfo {author} {\bibfnamefont
  {Y.}~\bibnamefont {Cai}}, \ and\ \bibinfo {author} {\bibfnamefont
  {O.}~\bibnamefont {Korotkova}},\ }\href@noop {} {\bibfield  {journal}
  {\bibinfo  {journal} {Physical Review A}\ }\textbf {\bibinfo {volume} {89}},\
  \bibinfo {pages} {013801} (\bibinfo {year} {2014})}\BibitemShut {NoStop}%
\bibitem [{\citenamefont {Partanen}\ \emph {et~al.}(2014)\citenamefont
  {Partanen}, \citenamefont {Turunen},\ and\ \citenamefont
  {Tervo}}]{Partanen:2014}%
  \BibitemOpen
  \bibfield  {author} {\bibinfo {author} {\bibfnamefont {H.}~\bibnamefont
  {Partanen}}, \bibinfo {author} {\bibfnamefont {J.}~\bibnamefont {Turunen}}, \
  and\ \bibinfo {author} {\bibfnamefont {J.}~\bibnamefont {Tervo}},\ }\href
  {\doibase 10.1364/OL.39.001034} {\bibfield  {journal} {\bibinfo  {journal}
  {Optics Letters}\ }\textbf {\bibinfo {volume} {39}},\ \bibinfo {pages} {1034}
  (\bibinfo {year} {2014})}\BibitemShut {NoStop}%
\bibitem [{\citenamefont {Pfeiffer}\ \emph {et~al.}(2005)\citenamefont
  {Pfeiffer}, \citenamefont {Bunk}, \citenamefont {Schulze-Briese},
  \citenamefont {Diaz}, \citenamefont {Weitkamp}, \citenamefont {David},
  \citenamefont {van~der Veen}, \citenamefont {Vartanyants},\ and\
  \citenamefont {Robinson}}]{Pfeiffer:2005}%
  \BibitemOpen
  \bibfield  {author} {\bibinfo {author} {\bibfnamefont {F.}~\bibnamefont
  {Pfeiffer}}, \bibinfo {author} {\bibfnamefont {O.}~\bibnamefont {Bunk}},
  \bibinfo {author} {\bibfnamefont {C.}~\bibnamefont {Schulze-Briese}},
  \bibinfo {author} {\bibfnamefont {A.}~\bibnamefont {Diaz}}, \bibinfo {author}
  {\bibfnamefont {T.}~\bibnamefont {Weitkamp}}, \bibinfo {author}
  {\bibfnamefont {C.}~\bibnamefont {David}}, \bibinfo {author} {\bibfnamefont
  {J.~F.}\ \bibnamefont {van~der Veen}}, \bibinfo {author} {\bibfnamefont
  {I.}~\bibnamefont {Vartanyants}}, \ and\ \bibinfo {author} {\bibfnamefont
  {I.~K.}\ \bibnamefont {Robinson}},\ }\href@noop {} {\bibfield  {journal}
  {\bibinfo  {journal} {Physical Review Letters}\ }\textbf {\bibinfo {volume}
  {94}},\ \bibinfo {pages} {164801} (\bibinfo {year} {2005})}\BibitemShut
  {NoStop}%
\bibitem [{\citenamefont {Divitt}\ and\ \citenamefont
  {Novotny}(2015)}]{Divitt:2015}%
  \BibitemOpen
  \bibfield  {author} {\bibinfo {author} {\bibfnamefont {S.}~\bibnamefont
  {Divitt}}\ and\ \bibinfo {author} {\bibfnamefont {L.}~\bibnamefont
  {Novotny}},\ }\href {\doibase 10.1364/OPTICA.2.000095} {\bibfield  {journal}
  {\bibinfo  {journal} {Optica}\ }\textbf {\bibinfo {volume} {2}},\ \bibinfo
  {pages} {95} (\bibinfo {year} {2015})}\BibitemShut {NoStop}%
\bibitem [{\citenamefont {Morrill}\ \emph {et~al.}(2016)\citenamefont
  {Morrill}, \citenamefont {Li},\ and\ \citenamefont
  {Pacifici}}]{Morrill:2016}%
  \BibitemOpen
  \bibfield  {author} {\bibinfo {author} {\bibfnamefont {D.}~\bibnamefont
  {Morrill}}, \bibinfo {author} {\bibfnamefont {D.}~\bibnamefont {Li}}, \ and\
  \bibinfo {author} {\bibfnamefont {D.}~\bibnamefont {Pacifici}},\ }\href@noop
  {} {\bibfield  {journal} {\bibinfo  {journal} {Nat Photon}\ }\textbf
  {\bibinfo {volume} {10}},\ \bibinfo {pages} {681} (\bibinfo {year}
  {2016})}\BibitemShut {NoStop}%
\bibitem [{\citenamefont {Marathe}\ \emph {et~al.}(2014)\citenamefont
  {Marathe}, \citenamefont {Shi}, \citenamefont {Wojcik}, \citenamefont
  {Kujala}, \citenamefont {Divan}, \citenamefont {Mancini}, \citenamefont
  {Macrander},\ and\ \citenamefont {Assoufid}}]{Marathe:2014}%
  \BibitemOpen
  \bibfield  {author} {\bibinfo {author} {\bibfnamefont {S.}~\bibnamefont
  {Marathe}}, \bibinfo {author} {\bibfnamefont {X.}~\bibnamefont {Shi}},
  \bibinfo {author} {\bibfnamefont {M.~J.}\ \bibnamefont {Wojcik}}, \bibinfo
  {author} {\bibfnamefont {N.~G.}\ \bibnamefont {Kujala}}, \bibinfo {author}
  {\bibfnamefont {R.}~\bibnamefont {Divan}}, \bibinfo {author} {\bibfnamefont
  {D.~C.}\ \bibnamefont {Mancini}}, \bibinfo {author} {\bibfnamefont {A.~T.}\
  \bibnamefont {Macrander}}, \ and\ \bibinfo {author} {\bibfnamefont
  {L.}~\bibnamefont {Assoufid}},\ }\href {\doibase 10.1364/OE.22.014041}
  {\bibfield  {journal} {\bibinfo  {journal} {Optics Express}\ }\textbf
  {\bibinfo {volume} {22}},\ \bibinfo {pages} {14041} (\bibinfo {year}
  {2014})}\BibitemShut {NoStop}%
\bibitem [{\citenamefont {Shi}\ \emph {et~al.}(2014)\citenamefont {Shi},
  \citenamefont {Marathe}, \citenamefont {Wojcik}, \citenamefont {Kujala},
  \citenamefont {Macrander},\ and\ \citenamefont {Assoufid}}]{Shi:2014}%
  \BibitemOpen
  \bibfield  {author} {\bibinfo {author} {\bibfnamefont {X.}~\bibnamefont
  {Shi}}, \bibinfo {author} {\bibfnamefont {S.}~\bibnamefont {Marathe}},
  \bibinfo {author} {\bibfnamefont {M.~J.}\ \bibnamefont {Wojcik}}, \bibinfo
  {author} {\bibfnamefont {N.~G.}\ \bibnamefont {Kujala}}, \bibinfo {author}
  {\bibfnamefont {A.~T.}\ \bibnamefont {Macrander}}, \ and\ \bibinfo {author}
  {\bibfnamefont {L.}~\bibnamefont {Assoufid}},\ }\href {\doibase
  10.1063/1.4892002} {\bibfield  {journal} {\bibinfo  {journal} {Applied
  Physics Letters}\ }\textbf {\bibinfo {volume} {105}},\ \bibinfo {pages}
  {041116} (\bibinfo {year} {2014})}\BibitemShut {NoStop}%
\bibitem [{\citenamefont {Tran}\ \emph {et~al.}(2007)\citenamefont {Tran},
  \citenamefont {Williams}, \citenamefont {Roberts}, \citenamefont {Flewett},
  \citenamefont {Peele}, \citenamefont {Paterson}, \citenamefont {de~Jonge},\
  and\ \citenamefont {Nugent}}]{Tran:2007}%
  \BibitemOpen
  \bibfield  {author} {\bibinfo {author} {\bibfnamefont {C.~Q.}\ \bibnamefont
  {Tran}}, \bibinfo {author} {\bibfnamefont {G.~J.}\ \bibnamefont {Williams}},
  \bibinfo {author} {\bibfnamefont {A.}~\bibnamefont {Roberts}}, \bibinfo
  {author} {\bibfnamefont {S.}~\bibnamefont {Flewett}}, \bibinfo {author}
  {\bibfnamefont {A.~G.}\ \bibnamefont {Peele}}, \bibinfo {author}
  {\bibfnamefont {D.}~\bibnamefont {Paterson}}, \bibinfo {author}
  {\bibfnamefont {M.~D.}\ \bibnamefont {de~Jonge}}, \ and\ \bibinfo {author}
  {\bibfnamefont {K.~A.}\ \bibnamefont {Nugent}},\ }\href@noop {} {\bibfield
  {journal} {\bibinfo  {journal} {Physical Review Letters}\ }\textbf {\bibinfo
  {volume} {98}},\ \bibinfo {pages} {224801} (\bibinfo {year}
  {2007})}\BibitemShut {NoStop}%
\bibitem [{\citenamefont {Waller}\ \emph {et~al.}(2012)\citenamefont {Waller},
  \citenamefont {Situ},\ and\ \citenamefont {Fleischer}}]{Waller:2012}%
  \BibitemOpen
  \bibfield  {author} {\bibinfo {author} {\bibfnamefont {L.}~\bibnamefont
  {Waller}}, \bibinfo {author} {\bibfnamefont {G.}~\bibnamefont {Situ}}, \ and\
  \bibinfo {author} {\bibfnamefont {J.~W.}\ \bibnamefont {Fleischer}},\
  }\href@noop {} {\bibfield  {journal} {\bibinfo  {journal} {Nat Photon}\
  }\textbf {\bibinfo {volume} {6}},\ \bibinfo {pages} {474} (\bibinfo {year}
  {2012})}\BibitemShut {NoStop}%
\bibitem [{\citenamefont {Wood}\ \emph {et~al.}(2014)\citenamefont {Wood},
  \citenamefont {Sharma}, \citenamefont {Cho}, \citenamefont {Brown},\ and\
  \citenamefont {Alonso}}]{Wood:2014}%
  \BibitemOpen
  \bibfield  {author} {\bibinfo {author} {\bibfnamefont {J.~K.}\ \bibnamefont
  {Wood}}, \bibinfo {author} {\bibfnamefont {K.~A.}\ \bibnamefont {Sharma}},
  \bibinfo {author} {\bibfnamefont {S.}~\bibnamefont {Cho}}, \bibinfo {author}
  {\bibfnamefont {T.~G.}\ \bibnamefont {Brown}}, \ and\ \bibinfo {author}
  {\bibfnamefont {M.~A.}\ \bibnamefont {Alonso}},\ }\href {\doibase
  10.1364/OL.39.004927} {\bibfield  {journal} {\bibinfo  {journal} {Optics
  Letters}\ }\textbf {\bibinfo {volume} {39}},\ \bibinfo {pages} {4927}
  (\bibinfo {year} {2014})}\BibitemShut {NoStop}%
\bibitem [{\citenamefont {Sharma}\ \emph {et~al.}(2016)\citenamefont {Sharma},
  \citenamefont {Brown},\ and\ \citenamefont {Alonso}}]{Sharma:2016}%
  \BibitemOpen
  \bibfield  {author} {\bibinfo {author} {\bibfnamefont {K.~A.}\ \bibnamefont
  {Sharma}}, \bibinfo {author} {\bibfnamefont {T.~G.}\ \bibnamefont {Brown}}, \
  and\ \bibinfo {author} {\bibfnamefont {M.~A.}\ \bibnamefont {Alonso}},\
  }\href {\doibase 10.1364/OE.24.016099} {\bibfield  {journal} {\bibinfo
  {journal} {Optics Express}\ }\textbf {\bibinfo {volume} {24}},\ \bibinfo
  {pages} {16099} (\bibinfo {year} {2016})}\BibitemShut {NoStop}%
\bibitem [{\citenamefont {Liu}\ \emph {et~al.}(2017)\citenamefont {Liu},
  \citenamefont {Wang}, \citenamefont {Liu}, \citenamefont {Chen},
  \citenamefont {Cai},\ and\ \citenamefont {Ponomarenko}}]{Liu:2017}%
  \BibitemOpen
  \bibfield  {author} {\bibinfo {author} {\bibfnamefont {X.}~\bibnamefont
  {Liu}}, \bibinfo {author} {\bibfnamefont {F.}~\bibnamefont {Wang}}, \bibinfo
  {author} {\bibfnamefont {L.}~\bibnamefont {Liu}}, \bibinfo {author}
  {\bibfnamefont {Y.}~\bibnamefont {Chen}}, \bibinfo {author} {\bibfnamefont
  {Y.}~\bibnamefont {Cai}}, \ and\ \bibinfo {author} {\bibfnamefont {S.~A.}\
  \bibnamefont {Ponomarenko}},\ }\href {\doibase 10.1364/OL.42.000077}
  {\bibfield  {journal} {\bibinfo  {journal} {Optics Letters}\ }\textbf
  {\bibinfo {volume} {42}},\ \bibinfo {pages} {77} (\bibinfo {year}
  {2017})}\BibitemShut {NoStop}%
\bibitem [{\citenamefont {Clark}\ \emph {et~al.}(2012)\citenamefont {Clark},
  \citenamefont {Huang}, \citenamefont {Harder},\ and\ \citenamefont
  {Robinson}}]{Clark:2012}%
  \BibitemOpen
  \bibfield  {author} {\bibinfo {author} {\bibfnamefont {J.~N.}\ \bibnamefont
  {Clark}}, \bibinfo {author} {\bibfnamefont {X.}~\bibnamefont {Huang}},
  \bibinfo {author} {\bibfnamefont {R.}~\bibnamefont {Harder}}, \ and\ \bibinfo
  {author} {\bibfnamefont {I.~K.}\ \bibnamefont {Robinson}},\ }\href@noop {}
  {\bibfield  {journal} {\bibinfo  {journal} {Nature Communications}\ }\textbf
  {\bibinfo {volume} {3}},\ \bibinfo {pages} {993} (\bibinfo {year}
  {2012})}\BibitemShut {NoStop}%
\bibitem [{\citenamefont {Burdet}\ \emph {et~al.}(2015)\citenamefont {Burdet},
  \citenamefont {Shi}, \citenamefont {Parks}, \citenamefont {Clark},
  \citenamefont {Huang}, \citenamefont {Kevan},\ and\ \citenamefont
  {Robinson}}]{Burdet:2015}%
  \BibitemOpen
  \bibfield  {author} {\bibinfo {author} {\bibfnamefont {N.}~\bibnamefont
  {Burdet}}, \bibinfo {author} {\bibfnamefont {X.}~\bibnamefont {Shi}},
  \bibinfo {author} {\bibfnamefont {D.}~\bibnamefont {Parks}}, \bibinfo
  {author} {\bibfnamefont {J.~N.}\ \bibnamefont {Clark}}, \bibinfo {author}
  {\bibfnamefont {X.}~\bibnamefont {Huang}}, \bibinfo {author} {\bibfnamefont
  {S.~D.}\ \bibnamefont {Kevan}}, \ and\ \bibinfo {author} {\bibfnamefont
  {I.~K.}\ \bibnamefont {Robinson}},\ }\href@noop {} {\bibfield  {journal}
  {\bibinfo  {journal} {Optics Express}\ }\textbf {\bibinfo {volume} {23}}
  (\bibinfo {year} {2015})}\BibitemShut {NoStop}%
\bibitem [{\citenamefont {Whitehead}\ \emph {et~al.}(2009)\citenamefont
  {Whitehead}, \citenamefont {Williams}, \citenamefont {Quiney}, \citenamefont
  {Vine}, \citenamefont {Dilanian}, \citenamefont {Flewett}, \citenamefont
  {Nugent}, \citenamefont {Peele}, \citenamefont {Balaur},\ and\ \citenamefont
  {McNulty}}]{Whitehead:2009}%
  \BibitemOpen
  \bibfield  {author} {\bibinfo {author} {\bibfnamefont {L.~W.}\ \bibnamefont
  {Whitehead}}, \bibinfo {author} {\bibfnamefont {G.~J.}\ \bibnamefont
  {Williams}}, \bibinfo {author} {\bibfnamefont {H.~M.}\ \bibnamefont
  {Quiney}}, \bibinfo {author} {\bibfnamefont {D.~J.}\ \bibnamefont {Vine}},
  \bibinfo {author} {\bibfnamefont {R.~A.}\ \bibnamefont {Dilanian}}, \bibinfo
  {author} {\bibfnamefont {S.}~\bibnamefont {Flewett}}, \bibinfo {author}
  {\bibfnamefont {K.~A.}\ \bibnamefont {Nugent}}, \bibinfo {author}
  {\bibfnamefont {A.~G.}\ \bibnamefont {Peele}}, \bibinfo {author}
  {\bibfnamefont {E.}~\bibnamefont {Balaur}}, \ and\ \bibinfo {author}
  {\bibfnamefont {I.}~\bibnamefont {McNulty}},\ }\href@noop {} {\bibfield
  {journal} {\bibinfo  {journal} {Physical Review Letters}\ }\textbf {\bibinfo
  {volume} {103}},\ \bibinfo {pages} {243902} (\bibinfo {year}
  {2009})}\BibitemShut {NoStop}%
\bibitem [{\citenamefont {Thibault}\ and\ \citenamefont
  {Menzel}(2013)}]{Thibault:2013}%
  \BibitemOpen
  \bibfield  {author} {\bibinfo {author} {\bibfnamefont {P.}~\bibnamefont
  {Thibault}}\ and\ \bibinfo {author} {\bibfnamefont {A.}~\bibnamefont
  {Menzel}},\ }\href@noop {} {\bibfield  {journal} {\bibinfo  {journal}
  {Nature}\ }\textbf {\bibinfo {volume} {494}},\ \bibinfo {pages} {68}
  (\bibinfo {year} {2013})}\BibitemShut {NoStop}%
\bibitem [{\citenamefont {McNulty}\ \emph {et~al.}(1992)\citenamefont
  {McNulty}, \citenamefont {Kirz}, \citenamefont {Jacobsen}, \citenamefont
  {Anderson}, \citenamefont {Howells},\ and\ \citenamefont
  {Kern}}]{McNulty:1992}%
  \BibitemOpen
  \bibfield  {author} {\bibinfo {author} {\bibfnamefont {I.}~\bibnamefont
  {McNulty}}, \bibinfo {author} {\bibfnamefont {J.}~\bibnamefont {Kirz}},
  \bibinfo {author} {\bibfnamefont {C.}~\bibnamefont {Jacobsen}}, \bibinfo
  {author} {\bibfnamefont {E.~H.}\ \bibnamefont {Anderson}}, \bibinfo {author}
  {\bibfnamefont {M.~R.}\ \bibnamefont {Howells}}, \ and\ \bibinfo {author}
  {\bibfnamefont {D.~P.}\ \bibnamefont {Kern}},\ }\href
  {http://science.sciencemag.org/content/256/5059/1009.abstract} {\bibfield
  {journal} {\bibinfo  {journal} {Science}\ }\textbf {\bibinfo {volume}
  {256}},\ \bibinfo {pages} {1009} (\bibinfo {year} {1992})}\BibitemShut
  {NoStop}%
\bibitem [{\citenamefont {Marchesini}\ \emph {et~al.}(2008)\citenamefont
  {Marchesini}, \citenamefont {Boutet}, \citenamefont {Sakdinawat},
  \citenamefont {Bogan}, \citenamefont {Bajt}, \citenamefont {Barty},
  \citenamefont {Chapman}, \citenamefont {Frank}, \citenamefont {Hau-Riege},
  \citenamefont {Szoke}, \citenamefont {Cui}, \citenamefont {Shapiro},
  \citenamefont {Howells}, \citenamefont {Spence}, \citenamefont {Shaevitz},
  \citenamefont {Lee}, \citenamefont {Hajdu},\ and\ \citenamefont
  {Seibert}}]{Marchesini:2008}%
  \BibitemOpen
  \bibfield  {author} {\bibinfo {author} {\bibfnamefont {S.}~\bibnamefont
  {Marchesini}}, \bibinfo {author} {\bibfnamefont {S.}~\bibnamefont {Boutet}},
  \bibinfo {author} {\bibfnamefont {A.~E.}\ \bibnamefont {Sakdinawat}},
  \bibinfo {author} {\bibfnamefont {M.~J.}\ \bibnamefont {Bogan}}, \bibinfo
  {author} {\bibfnamefont {S.}~\bibnamefont {Bajt}}, \bibinfo {author}
  {\bibfnamefont {A.}~\bibnamefont {Barty}}, \bibinfo {author} {\bibfnamefont
  {H.~N.}\ \bibnamefont {Chapman}}, \bibinfo {author} {\bibfnamefont
  {M.}~\bibnamefont {Frank}}, \bibinfo {author} {\bibfnamefont {S.~P.}\
  \bibnamefont {Hau-Riege}}, \bibinfo {author} {\bibfnamefont {A.}~\bibnamefont
  {Szoke}}, \bibinfo {author} {\bibfnamefont {C.}~\bibnamefont {Cui}}, \bibinfo
  {author} {\bibfnamefont {D.~A.}\ \bibnamefont {Shapiro}}, \bibinfo {author}
  {\bibfnamefont {M.~R.}\ \bibnamefont {Howells}}, \bibinfo {author}
  {\bibfnamefont {J.~C.~H.}\ \bibnamefont {Spence}}, \bibinfo {author}
  {\bibfnamefont {J.~W.}\ \bibnamefont {Shaevitz}}, \bibinfo {author}
  {\bibfnamefont {J.~Y.}\ \bibnamefont {Lee}}, \bibinfo {author} {\bibfnamefont
  {J.}~\bibnamefont {Hajdu}}, \ and\ \bibinfo {author} {\bibfnamefont {M.~M.}\
  \bibnamefont {Seibert}},\ }\href@noop {} {\bibfield  {journal} {\bibinfo
  {journal} {Nat Photon}\ }\textbf {\bibinfo {volume} {2}},\ \bibinfo {pages}
  {560} (\bibinfo {year} {2008})}\BibitemShut {NoStop}%
\bibitem [{\citenamefont {Codona}(2013)}]{Codona:2013}%
  \BibitemOpen
  \bibfield  {author} {\bibinfo {author} {\bibfnamefont {J.~L.}\ \bibnamefont
  {Codona}},\ }\href@noop {} {\bibfield  {journal} {\bibinfo  {journal}
  {Optical Engineering}\ }\textbf {\bibinfo {volume} {52}},\ \bibinfo {pages}
  {097105} (\bibinfo {year} {2013})}\BibitemShut {NoStop}%
\bibitem [{\citenamefont {Lai}\ \emph {et~al.}(2000)\citenamefont {Lai},
  \citenamefont {King},\ and\ \citenamefont {Neifeld}}]{Lai:2000}%
  \BibitemOpen
  \bibfield  {author} {\bibinfo {author} {\bibfnamefont {S.}~\bibnamefont
  {Lai}}, \bibinfo {author} {\bibfnamefont {B.}~\bibnamefont {King}}, \ and\
  \bibinfo {author} {\bibfnamefont {M.~A.}\ \bibnamefont {Neifeld}},\ }\href
  {\doibase http://dx.doi.org/10.1016/S0030-4018(99)00625-2} {\bibfield
  {journal} {\bibinfo  {journal} {Optics Communications}\ }\textbf {\bibinfo
  {volume} {173}},\ \bibinfo {pages} {155} (\bibinfo {year}
  {2000})}\BibitemShut {NoStop}%
\end{thebibliography}
%\bibliographystyle{plain}

\end{document}